\def\aa{A\&A}            
\def\anj{AJ}             
\def\apj{ApJ}            
\def\baas{BAAS}          
\def\mn{MNRAS}           
\documentclass{aa}

\begin{document}

\thesaurus{3(11.09.1 3C 395; 11.01.2; 11.10.1; 13.18.1)} 

\title{Space-VLBI observations of the twisted jet in 3C\,395}

\author{L. Lara\inst{1} \and 
A. Alberdi\inst{1} \and
J.M. Marcaide\inst{2} \and
T.W.B. Muxlow\inst{4}}

\offprints{L. Lara}

\institute{Instituto de Astrof\'{\i}sica de Andaluc\'{\i}a (CSIC),
Apdo. 3004, 18080 Granada (Spain)
\and 
Departamento de Astronom\'{\i}a, Universitat de Val\`encia, 46100 Burjassot - 
Spain
\and
NRAL, Jodrell Bank, Macclesfield, Cheshire SK11 9DL, UK
} 
\date{Received / Accepted}

\authorrunning{Lara et al.}
\titlerunning{Space-VLBI observations of 3C\,395}

\maketitle

\begin{abstract}

We present  Space-VLBI observations of  the quasar 3C\,395  which show
evidence of  the existence of a large  bend in the inner  parts of its
jet,  close  to  the  core.   This bend  would  explain  why  previous
Earth-based cm-VLBI observations could  not detect the ejection of new
moving components,  as expected from flux density  variability in this
source. In general, the observational properties of the jet in 3C\,395 
are heavily marked by the existence of bends on different scales.

\end{abstract}

\section{Introduction}

Flux density variability at radio  wavelengths in the compact cores of
radio loud  AGNs on time scales  of months or even  years is generally
associated  with ejection of  components in  parsec-scale relativistic
jets (e.g.  Valtaoja et al. \cite{valtaoja}).  Such  behavior has been
observed   in  many   radio   sources  (e.g.    Pauliny-Toth  et   al.
\cite{pauliny})  and is  successfully reproduced  by  theoretical work
(e.g.  Hughes et  al.  \cite{hughes}, G\'omez  et al.   \cite{gomez}).
However, there  are objects  which, apparently, do  not fit  into this
scenario.   One of  these,  the quasar  3C\,395 ($z=0.635$),  exhibits
significant flux density variability  which is clearly associated with
activity  in its  compact core  (Lara  et al.   \cite{lara1, lara2});  
however, Very Long  Baseline Interferometry (VLBI)
observations since 1984 show a stationary morphology, with no evidence
of the  expected correlation between flux density  variability and the
ejection of new jet components (Lara et al. \cite{lara2}).

In this paper we present new VLBI observations  of the quasar 3C\,395
at 4.8  GHz, made  with the  Very Long Baseline  Array (VLBA)  and the
Japanese satellite Halca.  These observations not only take profit of
the enhanced  angular resolution of Space-VLBI, but also  of the high
sensitivity achieved by the  continuous observation of a single source
at a  single frequency with the VLBA.  The new data shed  light on
the link  between flux density  variability and the structural properties 
of 3C\,395.

\section{Observations and data reduction}

\begin{figure}
\vspace{8cm}
\includegraphics{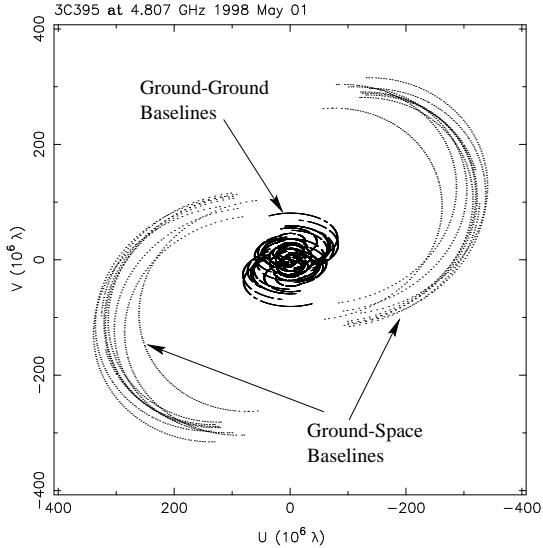}
\caption{uv-coverage obtained during the space-VLBI observations 
presented in this paper. The compact set of 
uv-data points corresponds to ground -- ground baselines. The outer 
tracks correspond to ground -- space baselines. The maximum uv-distance 
is 384 M$\lambda$.}
\label{fig1}
\end{figure}

We made continuum  observations of 3C\,395 with the  VLBA and Halca on
May 1st 1998  at a frequency of 4.8 GHz.   The observing bandwidth was
32 MHz.  Two tracking stations,  Robledo (Spain) and Green Bank (USA),
participated  in the  observations providing  maser  referenced timing
tones  to  the  satellite.  At   the  same  time,  they  received  the
astronomical  data from  Halca through  a Ku-band  downlink, recording
them on magnetic  tapes for later processing in  a VLBI correlator. In
Fig.~\ref{fig1}   we   display  the   uv-coverage   achieved  in   our
observations  to  illustrate  the  improvement in  angular  resolution
provided by  the orbiting  antenna.  The correlation  of the  data was
done {\em in absentia} by the  staff of the VLBA correlator in Socorro
(NM, USA). After correlation,  we used the NRAO
AIPS  package to  determine  the bandpass  response  functions of  the
antennas, to correct for  instrumental phase and delay offsets between
the  separate  baseband  converters  in  each  antenna,  to  determine
antenna-based  fringe corrections  and  to apply  the  {\em a  priori}
amplitude calibration.   Data imaging  in total intensity  was finally
performed with the Difmap package (Shepherd et al. \cite{shepherd}).

The imaging process consisted of two main steps: initially, we started
mapping the  VLBA data alone, i.e. without  ground-space baselines, at
low  angular resolution  by reducing  the  weight of  the longer  VLBA
baselines.   Once a  satisfactory low  resolution map  of  3C\,395 was
obtained, the weight of  the long baselines was progressively restored
to its  original values, so  that we finally  obtained a map  with the
VLBA alone  at its  maximum angular resolution.   In this  process, we
also  derived   accurate  self-calibration  solutions   for  the  VLBA
antennas.  We then included data  from Halca in a second mapping step,
improving  the  source  model  at  sub-milliarcsecond  resolution  and
obtaining a high  resolution map from the whole  data set. Finally, we
calibrated  the  absolute  flux  density  scale  mapping  the  compact
calibrator source  0133+476, also  observed during the  experiment and
with  an assumed  flux  density of  2.50  Jy at  4.8 GHz  (information
obtained from  the University of Michigan  Radio Astronomy Observatory
(UMRAO) database).

\section{Results}

\begin{figure*}
\vspace{13cm}
\includegraphics{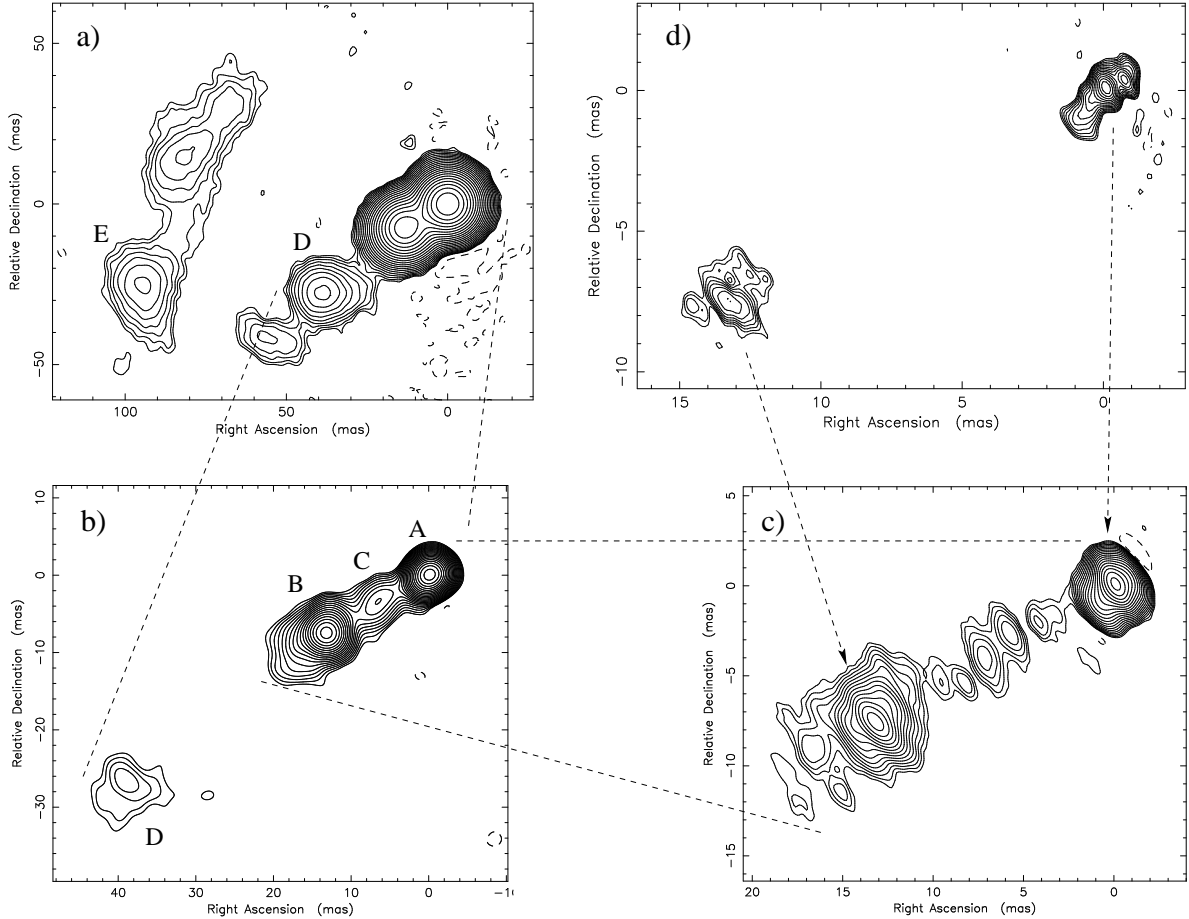}
\caption{{\bf a-d} VLBI maps of the jet in 3C\,395 at 4.8 GHz. {\bf a-b} 
correspond to the VLBA alone; {\bf c-d} correspond to the VLBA and Halca.
Dashed lines and arrows help to identify equivalent jet regions in the 
different maps. In all maps contours are spaced by factors of 
$\sqrt{2}$ in brightness, with the lowest at 3 times the rms noise level. 
For each map we list the Gaussian beam used in convolution (mas), the rms 
noise level 
(mJy\,beam$^{-1}$) and the peak
of brightness (Jy\,beam$^{-1}$). 
{\bf a:} Beam = 10$\times$10; rms = 0.17; Peak = 1.013;
{\bf b:} Beam = 2.5$\times$2.5; rms = 0.24; Peak = 0.894;
{\bf c:} Beam = 1.69$\times$0.76 P.A. 31.2$^{\circ}$; rms = 0.35; Peak = 0.608;
{\bf d:} Beam = 0.67$\times$0.31 P.A. 29.2$^{\circ}$; rms = 1.10; Peak = 0.415;
} 
\label{fig2}
\end{figure*}

Radio maps  of 3C\,395 are  displayed in Fig.~\ref{fig2}, with angular
resolutions spanning a range from 10 to 0.3 mas.

Fig.~\ref{fig2}a  shows a  low angular  resolution map  of the  jet in
3C\,395.  The brightness distribution  is dominated by a strong double
component, but the  most remarkable feature at this  resolution is the
existence of  a very sharp bend in  the jet at a  distance of $\sim$70
mas  from the core.  This bend  links the  compact structure  with the
extended emission observed at  sub-arcsecond resolution (Saikia et al.
\cite{saikia}; Lara et al. \cite{lara2}).  We can follow the jet up to
a length of nearly 200 mas  thanks to the very high sensitivity of the
VLBA.  The jet  at these large scales is not  smooth, but knotty.  Two
of these  knots can  be associated to  components D  and E in  Lara et
al.  (\cite{lara2}). The  total  flux density  recovered  in our  VLBI
observations is 1.43 Jy.

Fig.~\ref{fig2}b  has  an  angular   resolution  similar  to  that  of
previously published  cm-VLBI maps of 3C\,395 (Lara et
al. \cite{lara1}). Leaving  apart the weak
component D,  the radio structure can  be described in  terms of three
main  components named  A, B  and C,  as in  previous papers.  
Component A  has been usually identified as the radio
core;  component B  appeared  stationary  with respect  to  A and  was
interpreted  as the  result of  a  local bend  in the  jet towards  the
observer ;  component C,  between A  and B, was  claimed to  be moving
superluminally  after  VLBI  observations  during the  80's  (Waak  et
al. \cite{waak}; Simon et al. \cite{simon1}), but no clear evidence of
motion was  found in later  observations (Simon et  al. \cite{simon2};
Lara et al. \cite{lara2}).

Figs.~\ref{fig2}c-d reveal  that a  simple description of  the compact
structure of  3C\,395 in terms of  only three components  is no longer
valid when  observing with high sensitivity and  resolution. Feature A
is not  a single component: it  hosts the core, most  plausibly at its
western  edge, but  also  a  second strong  component  and a  jet-like
feature.  Such   complexity  within  A   was  suggested  by   Lara  et
al. (\cite{lara2}) after  a three-baseline  VLBI test observation  at 22
GHz.   Component B  also  shows  a rather  complex  structure at  high
angular    resolution.    Component    C    appears     resolved
(Fig.~\ref{fig2}c),  and it is difficult  to  associate it with any  single
feature, either moving or stationary.

To obtain a quantitative description of the milliarcsecond structure of
3C\,395, we have fitted simple elliptical Gaussian components to the visibility
data using a least square  algorithm within Difmap.  We needed a total
of  8 components  to satisfactorily  reproduce  the data;  
we did  not
attempt to  fit the brightness distribution beyond component D along  the 
jet, since this  extended emission   only  marginally affects the data  
from  the  shorter baselines. The  estimated parameters  for each 
Gaussian  component 
are given in Table~\ref{tab1}, where we  display the flux density (S), the
angular distance  from the westernmost component A1  (D), the position
angle with respect to A1 (P.A.), the length of the major axis (L), the
ratio between the major and minor  axis (r) and the orientation of the
major axis  ($\Phi$), defined in the  same sense as  the position angle.
The   elliptical   Gaussian    components   have   been   plotted   on
Fig.~\ref{fig4}  for  direct comparison  with  the surface  brightness
distribution.

\begin{table}[]
\caption[]{VLBI components of 3C\,395}
\label{tab1}
\begin{tabular}{ccccccc}
\hline
Comp.     &   S   &   D   &  P.A. &  L    & r & $\Phi$ \\
          & (mJy) & (mas) & (deg) & (mas) &   & (deg)  \\
\hline
A1 & 124  &    0     &   0    &   0.59   &  $<$0.01 &   106 \\
A2 & 629  &    0.66  &   109  &   0.30   &  0.74    &   132 \\
A3 & 245  &    1.57  &   126  &   0.58   &  0.71    &   160 \\
C  &  48  &    7.60  &   118  &   8.70   &  0.22    &   118 \\
B1 &  51  &   15.18  &   117  &   2.45   &  0.47    &   170 \\
B2 & 235  &   16.09  &   120  &   1.65   &  0.75    &    66 \\
B3 &  72  &   17.60  &   119  &   5.63   &  0.65    &   116 \\
D  &  16  &   48.53  &   125  &   7.39   &  0.80    &    53 \\
\hline
\end{tabular}
\end{table}

\section{Discussion and conclusions}

The launch of satellite Halca  in February 1997 has probably marked an
inflection  point  in  the   development  of  radio  astronomy.   VLBI,
continuously  improving  its  sensitivity  since its  origins  in  the
sixties, has finally managed to  break the hard constraints in angular
resolution imposed  by the limited size  of the Earth. As  a sample of
the  new  capabilities  of   VLBI,  our  observations  show  that  the
milliarcsecond  structure  of  3C\,395  is rather  more  complex  than
previously  thought,  with  several  features  which  deserve  special
attention.

First, there is a sudden decrease in the surface brightness of the jet
after component  A3, defining a sharp boundary  between this component
and the  rest of the jet.   Second, the jet position  angle between A2
and A3 is 137$^{\circ}$,  remarkably different from the position angle
defined by components  C and B (P.A. 118$^{\circ}$).   These two facts
argue towards the existence of a  bend in the jet soon after component
A3 in which  the orientation angle of the  jet increases significantly
with respect to the observer.  Third, as mentioned above, the emission
between components A  and B cannot be easily  associated with a single
component.  Nevertheless,  in our Gaussian  fit we have used  only one
component to  describe this emission  in order to compare  our results
with previous lower resolution  observations and look for any evidence
of motion,  but the  Gaussian component we  obtain is  very elongated,
making  any motion  estimates very  uncertain.  Moreover,  attempts to
model the complex structure of  C were not conclusive, suggesting that
the emission observed between components A and B might be related more
to the underlying jet hydrodynamics, rather than to a traveling 
shocked component.

\begin{figure}
\vspace{8cm}
\includegraphics{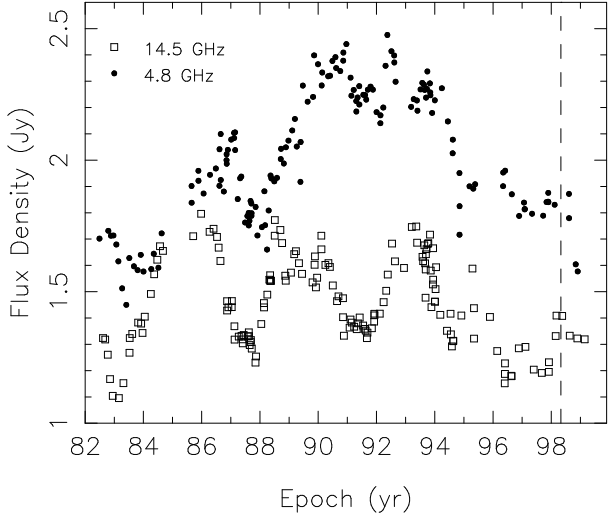} 
\caption{Time evolution of the flux density of 3C\,395 at 4.8 and 14.5 GHz,
obtained from  the UMRAO database. 8.4 GHz data, also available in this
database, have been excluded from this plot for the sake of clarity. The 
dashed line represents the date of the observations presented here.} 
\label{fig3}
\end{figure}

Another interesting  aspect of  3C\,395 is  related to  its variable flux
density. In Fig.~\ref{fig3}  we display the time evolution
in  the  flux density  of 3C\,395 at  4.8  and 14.5  GHz  since  1982. 
There  are
variations   of  up   to  30\%   in   total  flux   density  at   both
frequencies.  Monitoring  the  quasar with  VLBI  at  several
frequencies  and angular  resolutions since  1990 shows  that  the flux
density variability  is related with  the activity within  component A
(Lara et al. \cite{lara2}).

As noted before, flux density variability in the compact cores of AGNs
is  usually associated  to the  ejection of  new  traveling components
along  parsec-scale  relativistic   jets.   According  to  this  idea,
Fig.~\ref{fig3} suggests that new  components should have been ejected
from the  core of 3C\,395 in  1986, 1990 and  1993.  However, previous
VLBI  observations of  3C\,395 do  not show  the  expected correlation
between  flux  density  variations   and  the  ejection  of  new  VLBI
components.   The Space-VLBI  observations presented  here help  us to
understand this peculiarity:  if there exists a large  bend in the jet
after  component A3,  it  will  be very  difficult  to correlate  flux
density variability  within A  with structural variations  beyond this
component  because relativistic  time-delay makes  the time  scales of
these  two  events very  different.   Moreover,  the  decrease in  the
Doppler factor after  component A3 produces a large  diminution in the
flux  density  of  a  possible  moving  component.   3C\,395  requires
sub-milliarcsecond resolution to  study structural variations within A
associated  to   flux  density  variability.    Hence,  previous  VLBI
observations did not have the necessary angular resolution.

\begin{figure}
\vspace{8cm}
\includegraphics{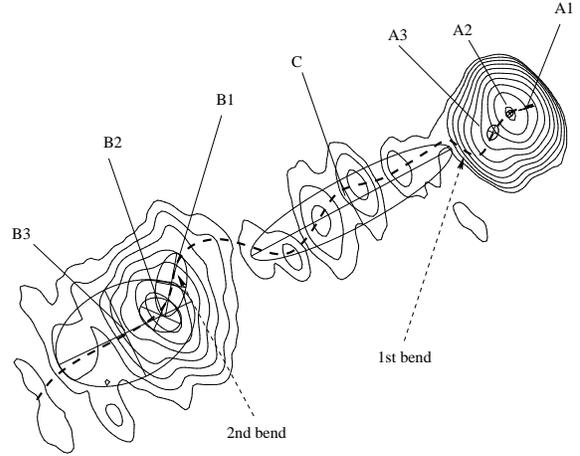} 
\caption{Space-VLBI image of 3C\,395 at 4.8 GHz with an approximate 
determination of the jet ridge line (dashed line). The positions of the 
bends which determine the observational properties of 3C\,395 are marked. 
The third sharp bend is out of this plot. Gaussian components in 
Table~\ref{tab1} are also marked, except component D which lies out of 
this plot.} 
\label{fig4}
\end{figure}

The ridge line  of the jet of 3C\,395 at  milliarcsecond scales can be
traced  from our maps  (see Fig.~\ref{fig4})  showing the  wiggles and
curvatures  present in  this twisted  jet.  We  can conclude  that the
observed properties of quasar 3C\,395 are heavily marked by three main
bends which produce changes in the orientation of the jet with respect
to the observer,  and hence changes in the  observed properties due to
Doppler  factor variations.   The first apparent bend, close  to the  
core and
producing a  departure of  the jet direction  from the line  of sight,
might be the reason why  flux density flares are not directly followed
by ejection of components  observable with ground cm-VLBI.  The second
bend,  at 15  mas from  the  core, orients  the jet  back towards  the
observer's  line   of  sight  and   may be  responsible of the  Doppler
amplification  of  the emission  at  this  position  resulting in  the
stationary component B.  The third bend,  at a distance of 70 mas from
the core, produces  a sharp deflection in the  projected ridge line of
the jet, appearing  to almost turn back upon  itself.  While the first
two  bends  could  possibly  be  consequence of  a  helically  twisted
geometry  in the  jet (which  might explain  also  components observed
beyond B), the third large curvature would require a global bending of
the helix. We note that the  intrinsic effects of these bends are most
probably highly amplified by a sharp overall orientation of the jet in
3C\,395 with respect to the observer's line of sight.

\begin{acknowledgements}

This   research  is   supported  in   part  by   the   Spanish  DGICYT
(PB97-1164). It has  made use of data from  the University of Michigan
Radio  Astronomy Observatory  which  is supported  by  funds from  the
University of Michigan. The  National Radio Astronomy Observatory is a
facility of the National Science Foundation operated under cooperative
agreement by Associated Universities, Inc.
\end{acknowledgements}

\end{document}